\documentclass[aps,prl,twocolumn,superscriptaddress]{revtex4-2}
\usepackage{multirow,amssymb,amsbsy,amsmath} 
\usepackage{graphicx} 
\usepackage{subfigure}
\usepackage[colorlinks,breaklinks]{hyperref}
\makeatletter
\newcommand\org@hypertarget{}
\let\org@hypertarget\hypertarget
\renewcommand\hypertarget[2]{%
\Hy@raisedlink{\org@hypertarget{#1}{}}#2% 
  }

\usepackage{xr}  
\usepackage{braket}
\makeatother 
\hypersetup{
colorlinks=true, % 启用颜色链接
linkcolor=blue, % 内部链接颜色
filecolor=blue, % 文件链接颜色
urlcolor=blue, % URL 链接颜色
citecolor=cyan % 引用链接颜色
}

\begin{document}
 
\title{Efficient integrated quantum memory based on a waveguide cavity
} 

\author{Pei-Xi Liu}
\author{Yi-Xin Xiao}
\author{Tian-Xiang Zhu}
\author{Zhe-Kai You}
\author{Zong-Quan Zhou}
\email{zq\_zhou@ustc.edu.cn}
\author{Chuan-Feng Li}
\author{Guang-Can Guo}

\affiliation{Laboratory of Quantum Information, University of Science and Technology of China, Hefei 230026, China}
\affiliation{Anhui Province Key Laboratory of Quantum Network, University of Science and Technology of China, Hefei 230026, China}
\affiliation{CAS Center For Excellence in Quantum Information and Quantum Physics, University of Science and Technology of China, Hefei 230026, China}
\affiliation{Hefei National Laboratory, University of Science and Technology of China, Hefei 230088, China}
\date{\today}

\begin{abstract}
Efficient integrated quantum memories are a key enabling technology for scalable quantum networks and photonic quantum processors, yet their efficiencies have remained below the critical threshold of 90\% required for demanding applications. Here we demonstrate an integrated photonic quantum memory with a storage-and-retrieval efficiency of 90.2(4)\% for single-photon-level weak coherent pulses and a storage fidelity of 99.3(1)\% for time-bin qubits, using a waveguide cavity fabricated in Pr$^{3+}$:$\mathrm{Y_2SiO_5}$. 
This performance is enabled by the high absorption-to-loss ratio of Pr$^{3+}$ ions in a low-loss waveguide cavity, combined with a time-reversal technique to overcome the dispersion-limited cavity linewidth and a novel spectral preparation scheme that creates a clean transparency window while utilizing a single class of ions that is compatible with spin-wave storage. Our results establish the first integrated photonic memory beyond the 90\% efficiency threshold, representing an important step toward applications that demand the highest performance, such as scalable quantum repeaters, photonic quantum computing and device-independent quantum key distribution.
\end{abstract}
\maketitle

Photonic quantum memories are fundamental building blocks for quantum repeaters and quantum networks~\cite{Kimble2008,RevModPhys.83.33,Ritter2012An,wehner2018quantum,Liu2021,doi:10.1126/science.1140300,yuan_experimental_2008,liu_creation_2024,zhu_metropolitan-scale_2026}, and also serve as enabling techniques for optical quantum computing~\cite{RevModPhys.79.135,PhysRevLett.110.133601,liu2024nonlocal} and long-baseline optical interferometry~\cite{stas_entanglement-assisted_2026,qpzn-h7p9}.
Memory efficiency is central to the overall performance of quantum repeaters. Even a slight reduction from 90\% to 89\% typically increases the distribution time by more than 10\%~\cite{RevModPhys.83.33}. Accordingly, a memory efficiency of 90\% has often been treated as a practically important benchmark for quantum repeater protocols~\cite{RevModPhys.83.33,PhysRevLett.113.053603,PhysRevLett.98.190503,PhysRevA.76.050301,PhysRevA.95.012319,krovi_practical_2016} and other advanced applications in quantum technologies ~\cite{PhysRevLett.112.250501,muralidharan_optimal_2016, PhysRevLett.98.230501, PhysRevLett.97.120501}. Although this level has been reached in cold atomic ensembles~\cite{Wang2019} and warm vapor cells~\cite{kbwj-md9n}, these systems are constrained by difficulties in miniaturization and integration. Consequently, an integrated photonic memory that achieves the 90\% efficiency level has, until now, remained an open challenge.

%due to their long coherence times~\cite{Zhong2015Optically,rancic_coherence_2018} and straightforward photonic integration~\cite{Zhou2023Photonic,8l9k-12k2,Saglamyurek2011,Zhong2017Nanophotonic}
Rare-earth-ion-doped crystals offer an attractive solid-state platform for integrated quantum memories ~\cite{Zhou2023Photonic,8l9k-12k2,Saglamyurek2011,Zhong2017Nanophotonic}, which provide unique advantages such as compact footprints, intrinsic mode confinement, stable alignment and low power consumption. However, achieving high efficiency on chip remains challenging because of the typically weak interaction between light and rare-earth ions. An impedance-matched cavity could in principle compensate for this limitation~\cite{PhysRevLett.110.133604}, but in practice the efficiency is often degraded by intra-cavity loss~\cite{8l9k-12k2}, imperfect mode matching~\cite{PhysRevLett.110.133604}, dispersion-limited bandwidth~\cite{Duranti2024Efficient}, and the complexity of bulk-cavity stabilization~\cite{8l9k-12k2}.

%. By coating mirrors on the waveguide facets, WGCs suppress mode mismatch and eliminate parasitic losses associated with free-space coupling, cryostat windows, and crystal surfaces.
The waveguide-based cavity (WGC)~\cite{Meng2026Efficient} provides a promising route to overcome these limitations by integrating microcavities directly into the waveguide platform, which has recently enabled a record storage efficiency of 80.3\% in solids, using Eu$^{3+}$:$\mathrm{Y_2SiO_5}$ crystals based on the atomic frequency comb (AFC) scheme. However, the absorption-enhanced pumping strategy employed to mitigate dispersion necessarily involves multiple classes of Eu$^{3+}$ ions in the AFC~\cite{Meng2026Efficient}, which inherently precludes direct extension to spin-wave storage~\cite{10.1093/nsr/nwae161,doi:10.1126/sciadv.adu5264,ortu_storage_2022,wvv1-6lg8}—a prerequisite for long-lived quantum memories.
%further improving the efficiency and reducing the device size remain great challenges due to the extremely weak absorption of Eu$^{3+}$. Moreover

Pr$^{3+}$:$\mathrm{Y_2SiO_5}$~\cite{Hedges2010,lago-rivera_telecom-heralded_2021,PhysRevLett.127.210502,wvv1-6lg8,z6lc-qw2d} offers significantly larger optical depth and cavity-enhanced quantum memories have been demonstrated in bulk Pr$^{3+}$:$\mathrm{Y_2SiO_5}$ crystals~\cite{PhysRevLett.110.133604,Duranti2024Efficient,8l9k-12k2}. However, the cavity-enhanced storage efficiency has so far been limited to 62\%~\cite{Duranti2024Efficient}, which remains below the 69\% efficiency achieved with a single-pass long-crystal configuration~\cite{Hedges2010}. This limitation primarily arises from large scattering losses in bulk cavities. Moreover, the large optical depth induces strong slow-light dispersion, further restricting storage efficiency through the associated narrow dispersion-limited bandwidth~\cite{Duranti2024Efficient}.

Here we overcome these obstacles and demonstrate an integrated photonic quantum memory using WGC based on Pr$^{3+}$:$\mathrm{Y_2SiO_5}$ with a storage efficiency of 90.2(4)\% for single-photon-level weak coherent pulses and an efficiency of 77.5(3)\% for time-bin qubits. Our approach combines three key innovations: a low-loss waveguide cavity with only 0.011 optical-depth round-trip loss, a spectral preparation scheme that isolates a single class of ions while creating a 16-MHz transparency window to suppress residual slow-light dispersion, and a time-reversal technique for optimizing the input waveform for the dispersion-limited storage bandwidth. % These results establish an integrated photonic quantum memory above the 90\% efficiency threshold and set a new benchmark for on-chip photonic quantum memories.Time-bin qubits are stored with an efficiency of 77.5(3)\% and an average fidelity of 99.3(1)\%, confirming the quantum nature of the memory. 

\begin{figure}[t]
\centering 
\includegraphics[width=1 \linewidth]{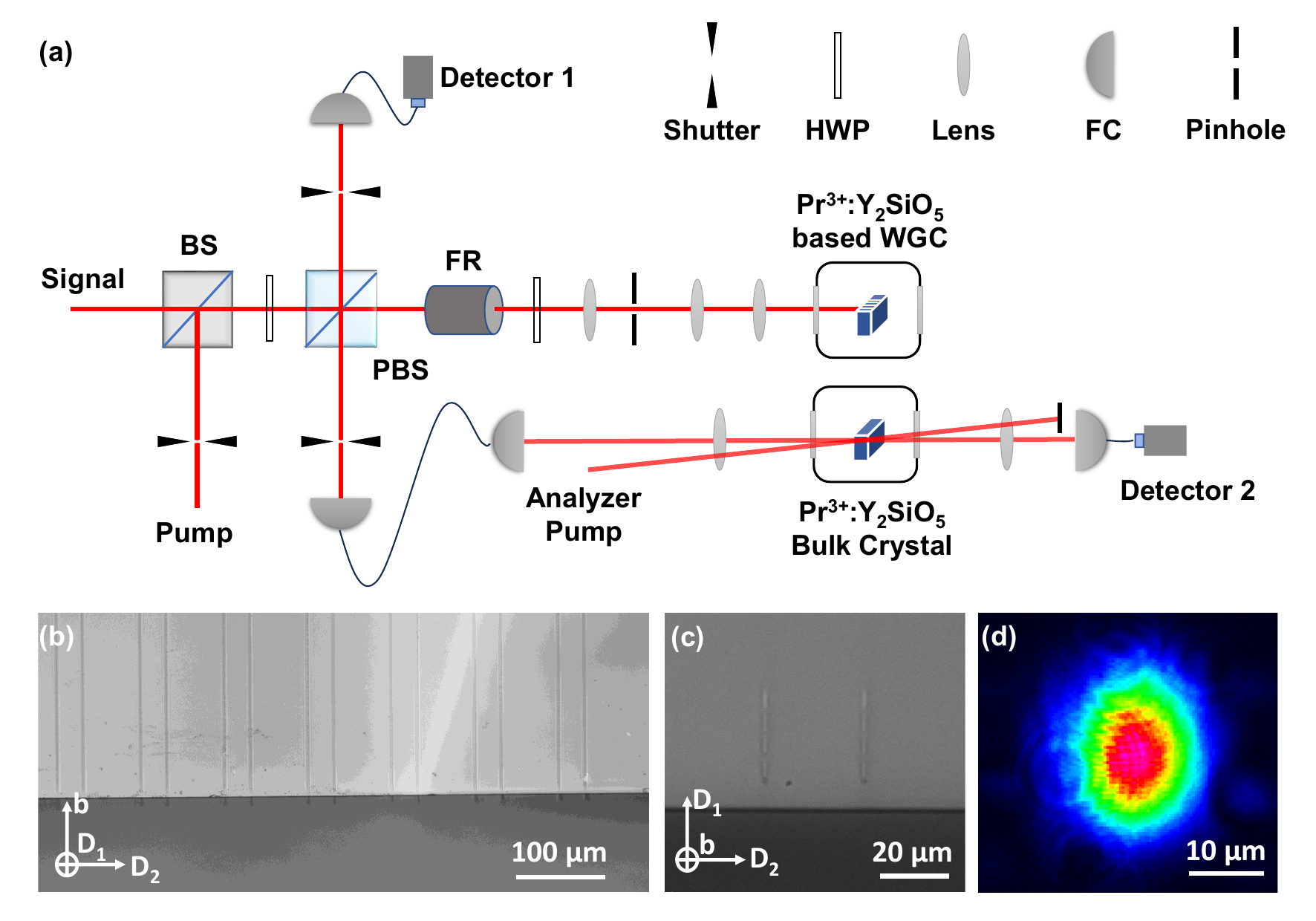}
    \caption{Experimental setup and device characterization.
    {(a)} Schematic of the experimental setup. 
    The signal and pump beams are combined using a beam splitter (BS) and coupled into the cavity mode using a lens group. The small transmission from the cavity is used as a criterion for optimizing the coupling. The retrieved echo is detected by a silicon-based single-photon detector (Si-SPD, Detector 2), and another Si-SPD (Detector 1) is used as a reference before the storage process, separated by a polarization beam splitter (PBS) and a Faraday rotator (FR). Another Pr$^{3+}$:$\mathrm {Y_2SiO_5}$ crystal is employed for analysis of superposition time-bin qubits. Both WGC and analyzer crystal are cooled to 3.2 K in cryostats.
    Si-SPDs are protected by shutters. Pinholes are employed to spatially reject spurious light. HWP: half-wave plate. FC: fiber coupler. 
    {(b)} Top view of the waveguides array. {(c)} Front view of the WGC. {(d)} Beam profile of the guided mode in the waveguide cavity.
}      
\label{setup}
\end{figure} 

%In this work, we overcome these challenges and demonstrate an integrated photonic quantum memory on Pr$^{3+}$:$\mathrm {Y_2SiO_5}$ platform with record-high efficiency. Our advances are threefold. First, we optimize the waveguides to achieve an ultra-low intra-cavity loss of 0.011 (in OD) per round trip. Second, we propose a novel spectral preparation scheme that preserves a single class of ions while creating a 16 MHz transparency pit, thereby minimizing residual slow light dispersion. Third, we introduce the time reversal to the waveform of input pulses to match the finite cavity linewidth. Through these innovations, we achieve a storage-retrieval efficiency of 90.2(4)\% for single-photon-level weak coherent pulses at a storage time of 1.8 \textmu s. We further confirm the quantum property of the memory by storing time-bin qubits with an average efficiency of 77.5(3)\% and an average fidelity of 99.3(1)\%. This work represents the first integrated quantum memory to exceed the 90\% efficiency threshold, establishing a new performance benchmark for on-chip photonic quantum storage and paving the way toward scalable integrated quantum networks and processors.

\begin{figure}[t]
\centering 
\includegraphics[width=1 \linewidth]{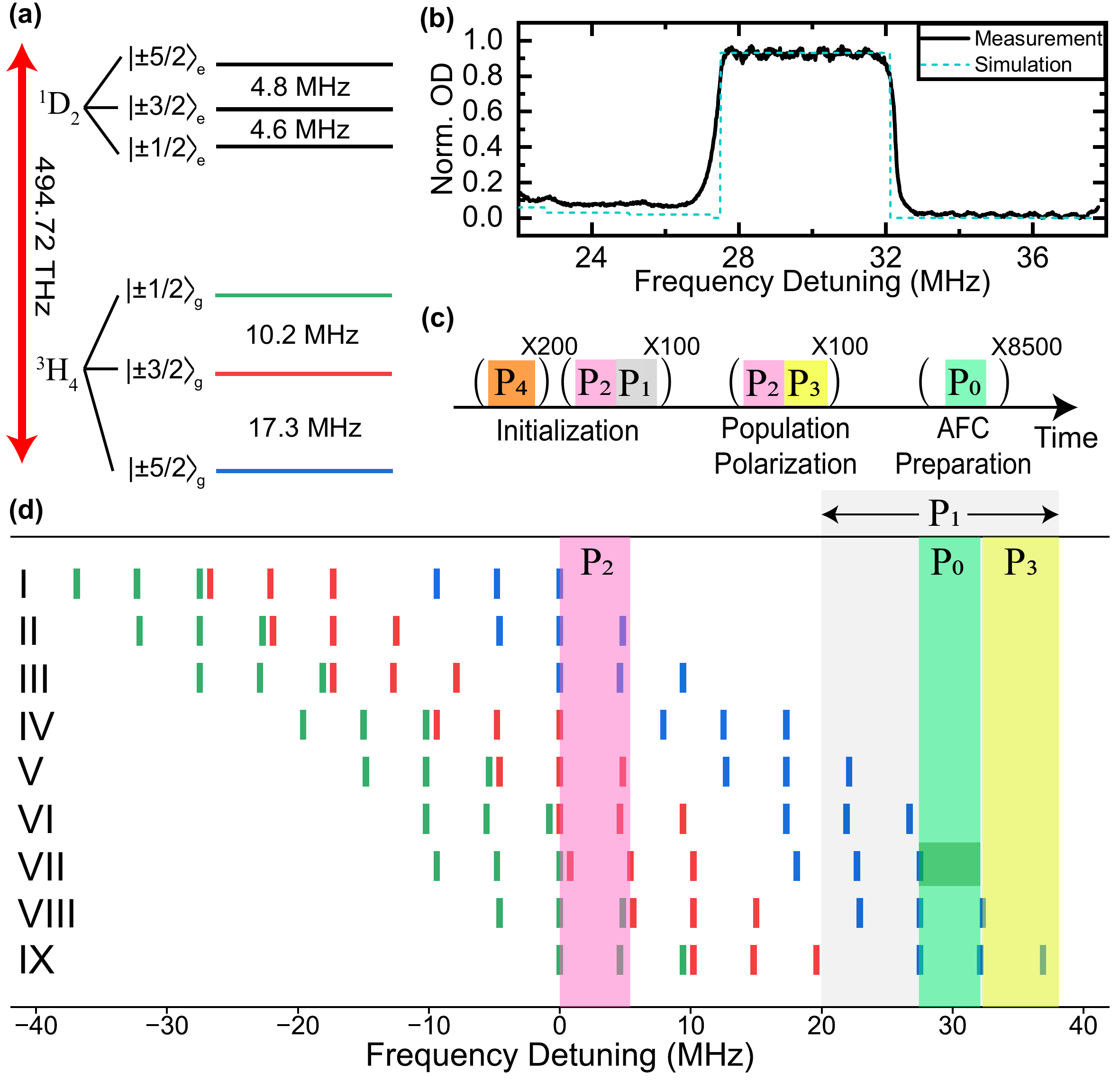}
    \caption{{(a)} The hyperfine level structure of the $\mathrm{{^3}H{_4}\rightarrow{^1}D{_2}}$ transition for site-1 $\mathrm{Pr^{3+}}$ ions in $\mathrm{Y_2SiO_5}$ crystal at the geomagnetic field. 
    {(b)} Normalized optical depth (OD) of the target absorption band after the population polarization process of the pumping scheme. The measured results (black solid line) are obtained with a bulk crystal without a cavity, and the simulated results are shown with a cyan dotted line.
    {(c)} The pulse sequence for the pump scheme consists of initialization, population polarization, and AFC preparation. Each pulse is represented by a rectangular bar colored according to its frequency range, with the number above the bracket indicating the repetition count. The frequency ranges of the pulses are denoted as follows: $\mathrm{P}_{0}$ (covering 27.5 to 32.1 MHz) marked in green, $\mathrm{P}_{1}$ (covering 20.0 to 38.0 MHz) in gray, $\mathrm{P}_{2}$ (covering 0 to 5.4 MHz) in pink, $\mathrm{P}_{3}$ (covering 32.3 to 38.0 MHz) in yellow, and $\mathrm{P}_{4}$ (covering -27.9 to 92.1 MHz) in orange, respectively. $\mathrm{P}_{4}$, $\mathrm{P}_{1}$, $\mathrm{P}_{2}$, and $\mathrm{P}_{3}$ are pumped during the initialization and population polarization, while $\mathrm{P}_{0}$ denotes the target frequency bandwidth. 
    {(d)} Diagram of the hyperfine transition arrangement for the 9-classes $\mathrm{Pr^{3+}}$ ions. 
    The 9-classes $\mathrm{Pr^{3+}}$ ions (I-IX, shown top to bottom) are defined by the transitions that occur at zero incident light detuning. Transitions originating from the ground states ($|\pm1 /2\rangle_g$, $|\pm3/2\rangle_g$, and $|\pm5/2\rangle_g$) are depicted by green, red, and blue lines, respectively \cite{Zhu2022On-Demand,Meng2026Efficient}.
    The frequency range $\mathrm{P}_{0}$, $\mathrm{P}_{1}$, $\mathrm{P}_{2}$, and $\mathrm{P}_{3}$ are indicated in {(c)}.
    After applying our pumping scheme, only a single class of $\mathrm{Pr^{3+}}$ ions (Class VII, in the $|\pm5/2\rangle_g$ state) remains populated within this bandwidth, indicated by the shaded area in $\mathrm{P}_{0}$.
}      
\label{pump_scheme}
\end{figure} 

% Memory efficiency drops as the input pulse bandwidth increases, reflecting the narrow cavity linewidth induced by slow-light dispersion from the high natural absorption of Pr$^{3+}$:$\mathrm {Y_2SiO_5}$.  Higher order echoes ($\sim1.0\%$ of the input) beyond 3 \textmu s are omitted for clarity. 
\begin{figure*}[t]
\centering 
\includegraphics[width=0.9 \linewidth]{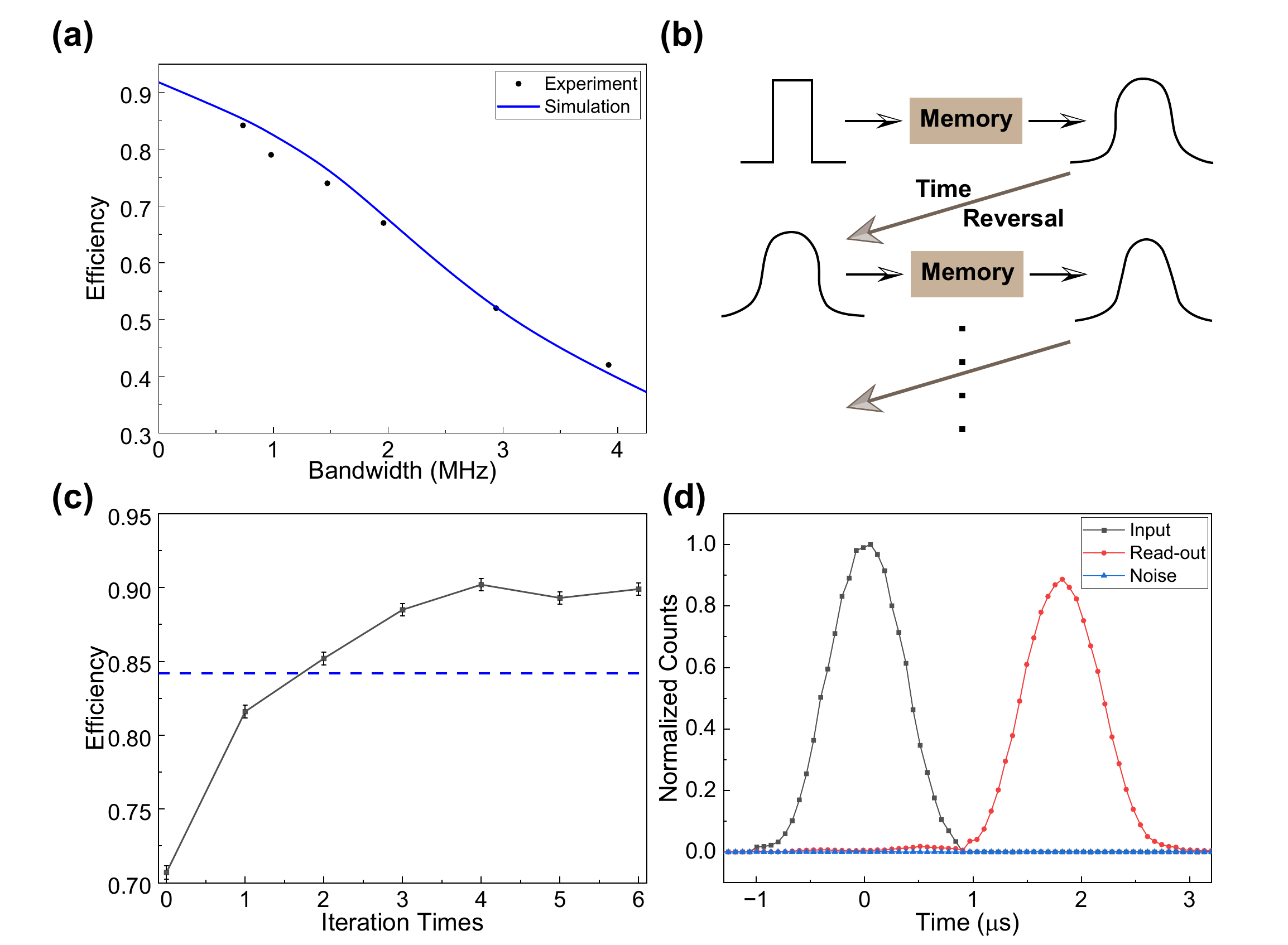}
    \caption{Efficient storage of single-photon level inputs. 
    {(a)} Measured (black circle) and simulated (blue line) efficiency as a function of bandwidth of Gaussian-shaped input pulse. Experimental points are measured using different input pulse bandwidth and the same AFC structure as used in {(d)}. Simulation method is detailed in Supplementary Materials Section D.
    {(b)} Illustration of time-reversal method. We start from a square shaped input pulse to show the robustness of time-reversal to the shape of input.
    {(c)} Evolution of efficiency with successive time-reversal iterations of the input waveform (red solid line). The blue dashed line marks the efficiency obtained with standard Gaussian shape input, corresponding to the highest measured efficiency in {(a)}. After two reversals, the efficiency exceeds the Gaussian-pulse baseline, and saturates after four times of reversal. The slight efficiency fluctuations are attributed to measurement and modulation imperfections. 
    {(d)} Normalized photon counting histograms for the AFC echo of 90.2(4)\% efficiency. The red line with solid circles represents the retrieved echo, at an average photon number per pulse $\mu=0.6$. The black line with solid squares is the reference input signal. Noise measured with the input blocked is indicated by the blue line with solid triangle.
}      
\label{efficiency}
\end{figure*} 

%Following the approach in Ref.~\cite{Meng2026Efficient}, we use coated crystals to accurately characterize the waveguide propagation loss by measuring the resonance reflection.
\textit{WGC fabrication and experimental setup.}
We use a 500 ppm doped Pr$^{3+}$:$\mathrm {Y_2SiO_5}$ crystal as the substrate. The crystal dimensions are $2.5\times6\times1.5$ mm$^3$ along $D_1$, $D_2$ and $b$ axes, with a 10 arcsec wedge polished between the front and end surfaces ($D_1-D_2$ surfaces) to facilitate cavity resonance tuning~\cite{PhysRevLett.110.133604}. An array of type-IV femtosecond laser micromachining (FLM) waveguides are fabricated 21 \textmu m below the top surface ($D_2-b$ surface), with a spacing of 100 \textmu m apart along the $D_2$ axis, as illustrated in Fig.~\ref{setup} (b-c). The waveguide mode is $\mathrm{TE_{00}}$, propagating along $b$ axis with polarization along $D_2$ (Fig.~\ref{setup}(d)). The optimal waveguide design consists of four notches on each side with a notch separation of 28 \textmu m, resulting in a single-pass waveguide loss corresponding to an optical depth of 0.005 (see Supplementary Materials~\cite{SupplementalMaterial,Seri:18,Rakonjac2022Storage,Guillot2009Hyperfine,Taherizadegan:23,POLLNAU2020100255,Gundogan2015Solid,Marcos2005Intercept-resend} Section A for details of characterization and optimization). The cavity is formed by mirrors coated on front and end surface with $R_{1}=0.7$ and $R_{2}=0.9988$, and is mounted inside a cryostat operating at 3.2 K. A schematic of the experimental setup is shown in Fig.~\ref{setup}(a). To accurately normalize the retrieved echo signal, an optically thick spectral feature is prepared in the crystal, and the input signal is determined by dividing the reflection pulse of the blocked cavity by $R_1$~\cite{Duranti2024Efficient}. The input is simultaneously monitored with a separate silicon-based single-photon detector (Si-SPD, Detector 1).
%The pulse energy used for fabricating the notches are 63.1 nJ, 61.3 nJ, 59.5 nJ and 58.0 nJ for notches from bottom to top

%Specifically, our sample has the absorption center at $494.7274$ THz with a single-pass optical depth of $d=7.1$ in the waveguides, leading to a 0.4 MHz cavity linewidth in a 16 MHz spectral pit~\cite{Sabooni_2013}. Our sample has the absorption center at $494.7274$ THz with a single-pass optical depth of $d=7.1$ in the waveguides.
\textit{Pumping scheme and AFC preparation.}
Pr$^{3+}$ ions in $\mathrm {Y_2SiO_5}$ crystal have relatively strong absorption at the optical transition of $\mathrm{{^3}H{_4}\rightarrow{^1}D{_2}}$, which leads to strong dispersion that limits the cavity linewidth (Fig.~\ref{pump_scheme}{(a)}). To suppress the slow-light effect induced by spectral hole burning, we set the laser frequency to $494.7248$ THz (2.6 GHz red-detuned from the sample absorption center), where the natural optical depth is $d = 4.0$, and design a new pumping scheme (Fig.~\ref{pump_scheme}(c)) to deplete the population as much as possible over a broad bandwidth around the absorption band.
The transition frequency diagram for the 9 classes of Pr$^{3+}$ ions is shown in Fig.~\ref{pump_scheme}{(d)}, where we select a specific frequency on the absorption band as the zero-detuning reference. 
The first step includes a 120-MHz-wide broadband frequency sweep centered on the target band $\mathrm P_{0}$ for initialization (green shaded area in Fig.~\ref{pump_scheme}(d)), and additional pumping at $\mathrm P_{1}$ and $\mathrm P_{2}$ frequency bands (marked in Fig.~\ref{pump_scheme}{(d)}) to deeply clear the excess population from other classes of ions. 
The second step is to simultaneously pump two frequency bands ($\mathrm P_{2}$ and $\mathrm P_{3}$), where pumping $\mathrm P_{2}$ brings the population back into the target bandwidth, and pumping $\mathrm P_{3}$ cleans other ion classes and residual population within the bandwidth. 
All steps are implemented using a series of weak, linearly frequency-swept pulses with 1 ms duration and repeat times shown in Fig.~\ref{pump_scheme}{(c)}. 
To verify the resulting absorption profile, we measure the sample absorption with this pumping scheme using another bulk crystal without a cavity. Fig.~\ref{pump_scheme}(b) shows that this scheme creates a 4.6-MHz flat absorption band and nearly fully depletes the population over a 16-MHz surrounding range, in good agreement with the simulated profile (see Supplementary Materials Section B for details).
After this, we create the AFC structure using chirped hyperbolic secant (CHS) pulses~\cite{liu2024nonlocal} with 4 MHz bandwidth, 1 ms duration, and 8500 repeating times. When the impedance-match condition is satisfied, the AFC finesse $F_{\mathrm {AFC}}$ is approximately 21, which leads to a high AFC rephasing efficiency. Most importantly, this pumping scheme directly supports spin-wave quantum storage by selecting a single class of $\mathrm{Pr^{3+}}$ ions that resonates with the $|\pm5/2\rangle_g \rightarrow |\pm5/2\rangle_e$ transition. For future spin-wave storage, a control pulse resonant with the $|\pm1/2\rangle_g \rightarrow |\pm5/2\rangle_e$ transition could map the optical coherence onto the spin coherence between the $|\pm1/2\rangle_g$ and $|\pm5/2\rangle_g$ states. Although this control transition has a relative oscillator strength of only 0.07~\cite{PhysRevB.70.214116}, the tight optical confinement of the waveguide should keep the required control-pulse peak power in the milliwatt range~\cite{PhysRevApplied.5.054013}. Moreover, the RF power required for spin manipulation can also be substantially reduced by using integrated coplanar waveguides~\cite{doi:10.1126/sciadv.adu5264}.
%Based on the transition branching ratios of Pr$^{3+}$ ions \cite{PhysRevB.70.214116} and assuming a completely uniform initial distribution of the nine classes of ions
%Consequently, our scheme offers a viable pathway toward long-lived, high-efficiency integrated quantum memories for practical applications.
%This pumping strategy is different from what we used in Ref. ~\cite{Meng2026Efficient}, where multiple class of $\mathrm{Eu^{3+}}$ ions are employed to create an enhanced absorption.

The dispersion of the prepared absorption profile without AFCs can be analyzed using the method detailed in Supplementary Materials Section C. Using the natural absorption coefficient $\alpha~(\mathrm{cm}^{-1})$ as a common normalization, we estimate that the pumping scheme employed in our experiment yields a slow-light contribution of $v_p/v_g = 180\alpha$ at the memory frequency, where $v_p$ and $v_g$ are the phase and group velocities, respectively. Applying the same analysis to the pumping scheme of Ref.~\cite{8l9k-12k2} yields $v_p/v_g = 280\alpha$, implying that our pumping approach suppresses the slow-light-induced dispersion by roughly 36\%, corresponding to an effective cavity linewidth enhancement from 0.43~MHz to 0.67~MHz. For an input pulse with FWHM = 0.6 $\mu$s, the simulated storage efficiencies are 74\% and 84\% for the two pumping schemes, respectively, reflecting a 10\% enhancement in efficiency as a result of dispersion minimization.%This corresponds to an effective cavity linewidth enhancement from 0.43~MHz to 0.67~MHz. After AFC structure prepared, about $\frac{\widetilde{d}\Gamma_{pit}}{d_{\mathrm{pit}}\Gamma_{AFC}}\approx18\%$ of dispersion will be neutralized~\cite{Meng2026Efficient}, where $\widetilde{d}$ is the effective optical depth, making the linewidth a little broader.

\textit{Time-reversal of the input waveform.}
After the preparation of AFCs with comb separation $\Delta=0.625$ MHz, we observe a retrieved echo after 1.8 \textmu s (including a 0.2 \textmu s delay due to slow light effect) with 84.2\% efficiency for a Gaussian-shaped input pulse, which is in accordance with our simulation (the first experiment point of Fig.~\ref{efficiency}(a)). However, as shown in Fig.~\ref{efficiency}(a), the simulated efficiency reaches 91.8\% under the narrow bandwidth limit, which suggests residual dispersion still limits the memory efficiency and the shape of the input pulse should be optimized for the limited storage bandwidth. 

%By shaping the input pulse into the time-reversed waveform of the output, the storage-and-retrieval efficiency can be significantly enhanced.
%The present cavity-coupled memory system is a typical bandwidth-limited system: the residual slow-light dispersion narrows the effective memory bandwidth, and thus the incident temporal mode must be matched to the dispersion-limited response.
Novikova \textit{et al.} developed a universal theory~\cite{novikova_optimal_2007,gorshkov_universal_2007} for quantum memories based on $\Lambda$ system in atomic ensembles,  demonstrating that the optimal retrieval process corresponds to the time reversal of the storage process. This approach has been successfully demonstrated in electromagnetically induced transparency (EIT) protocol~\cite{PhysRevA.78.021802,PhysRevLett.110.083601,Zhou:12,ma_high-performance_2022} and empty cavities~\cite{Bader_2013}.
The present cavity-coupled AFC memory, which can be regarded as a dispersive medium with spectral modulation~\cite{Meng2026Efficient}, exhibits the same time-reversal symmetry between storage and retrieval processes. 
We therefore adopt this time-reversal strategy to optimize the input pulse waveform to further improve the storage efficiency. We begin with a square wave pulse with a duration of 1.6 \textmu s, and record its retrieved echo using a Si-SPD. As we use a double-pass acousto-optic modulator (AOM) for pulse generation, where the output pulse intensity scales as the fourth power of the applied RF voltage, the fourth root of the time-reversed echo's photon-counting histogram is then used as the electric pulse waveform profile (Fig.~\ref{efficiency}(b)). After four iterations of this reversal–feedback procedure, the efficiency stabilizes around 90\% (Fig.~\ref{efficiency}(c)).
%To ensure the input faithfully reproduces the reversed echo, the waveform is multiplied by a linear correction factor that compensates for the AOM's nonlinear response. The full width at half maximum (FHWM) of the optimal input pulse is 0.82 \textmu s. 

With this optimized input, we achieve a storage-and-retrieval efficiency of 90.2(4)\% at an average photon number per pulse $\mu=0.6$ (Fig.~\ref{efficiency}(d)). 
Given an AFC finesse $F_{\mathrm{AFC}}=21$ and round-trip loss $\epsilon=0.011$, our result is very close to the theoretically predicted efficiency of 92\% using Eq.~1 in Ref.~\cite{Meng2026Efficient} and simulated maximum efficiency 91.8\% under the narrow-linewidth limit (detailed in Supplementary Materials Section D).
Only 1.7\% reflection and 1.0\% higher-order echoes remain, which indicates a nearly perfect spectral match between the input and cavity.

\begin{figure}[t]
\centering 
\includegraphics[width=1 \linewidth]{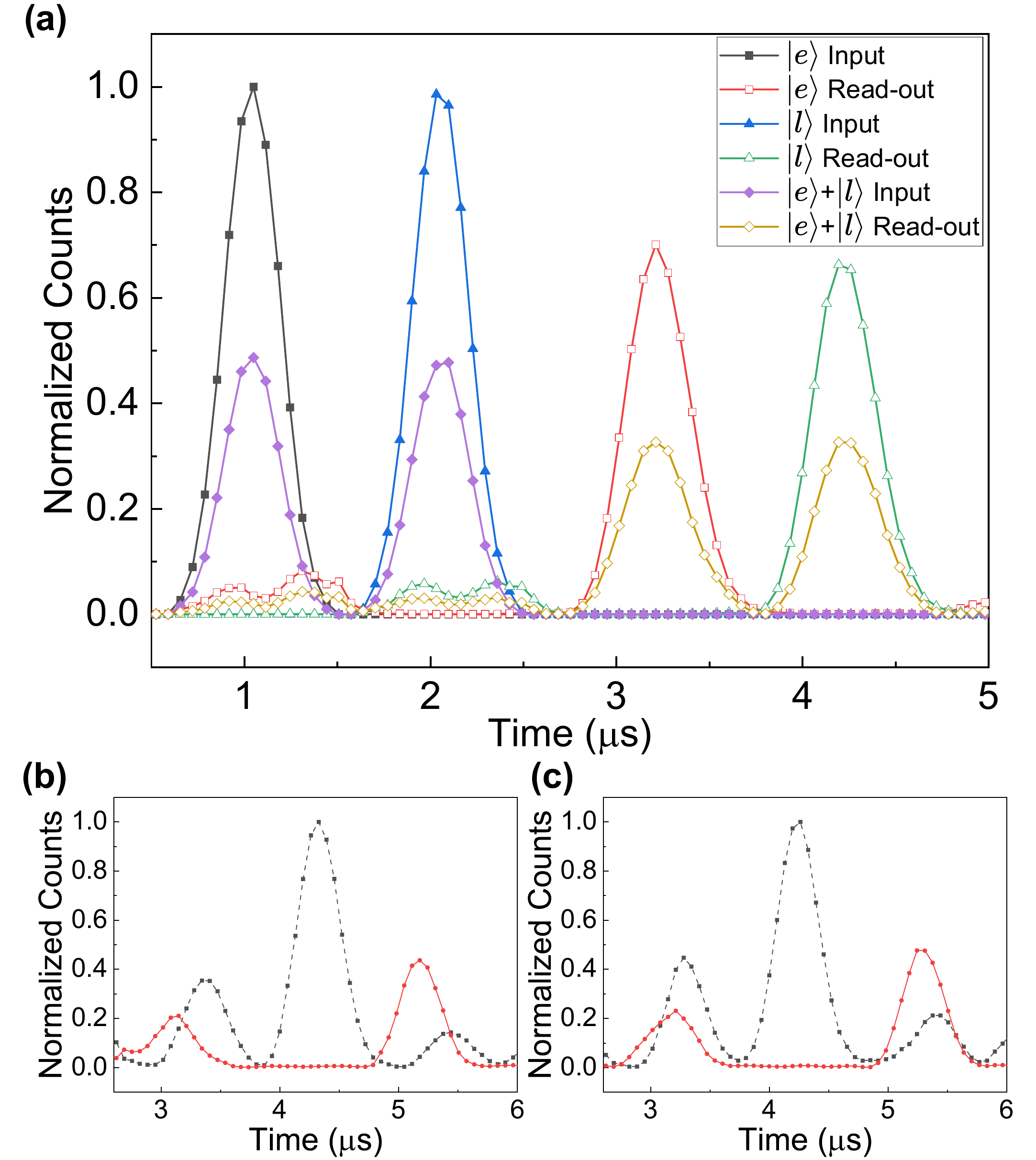}
    \caption{Efficient storage of time-bin qubits. {(a)}. Histograms for the storage of $\ket{e},\ket{l}$ and $\ket{e}+\ket{l}$. The average photon number per qubit $\mu_q=0.6$, with an average storage efficiency of 77.5(3)\%. {(b-c)}. Interference measurements on the superposition state $\ket{e}+\ket{l}$ and $\ket{e}+i\ket{l}$, respectively. Black dashed lines with squares represent the constructive interference, while red solid lines with circles represent the destructive interference.
}      
\label{timebin}
\end{figure}

\textit{Storage of time-bin qubits.}
To evaluate the quantum storage fidelity, we encode time-bin qubits onto the input photons. The FWHM of the Gaussian-shaped pulse states $\ket{e}$ and $\ket{l}$ is 0.36 \textmu s. The storage results of $\ket{e},\ket{l}$ and $\ket{e}+\ket{l}$ states are presented in Fig.~\ref{timebin}(a). The average efficiency for storage of the two-mode qubits is 77.5(3)\%. This reduction in efficiency compared to single-mode storage is attributed to the dispersion-limited cavity linewidth of 0.67 MHz. Nonetheless, it sets a new record for photonic qubit storage efficiency in solids (compared to the previous record 74.6\%~\cite{Meng2026Efficient}). The time-reversal method is not employed here, as the retrieved echo becomes broader after reversal and overlaps with itself. Extending the separation between the two temporal bins would in principle enable the time-reversal method, although the storage time is currently constrained by the limited finesse of spectral tailoring due to the relatively broad laser linewidth (FWHM over 5 kHz). Future advances in preparing AFCs with longer storage time~\cite{liu2024nonlocal} or employing other qubit encoding strategies with more memory channels~\cite{Wang2019,OZW}, should allow qubit storage efficiency to reach the same level as single-mode storage.
The fidelities of the superposition states are characterized using another AFC memory that serves as an unbalanced Mach–Zehnder interferometer~\cite{Duranti2024Efficient,Meng2026Efficient}, with the results summarized in Table~\ref{fidelity} (Supplementary Materials Section E). The overall storage fidelity exceeds 99\%, far above the classical bound (70.0\%) achievable with a measure-and-prepare strategy by taking into account the photon statistics and finite storage efficiency~\cite{specht_single-atom_2011,Meng2026Efficient}, unambiguously certifying that the memory operates in the quantum regime.

\begin{table}
    \begin{ruledtabular}
    \begin{tabular}{cccccc}

        $F_{|e\rangle}$ & 
        $F_{|l\rangle}$ & 
        $F_{|e\rangle+|l\rangle}$ & 
        $F_{|e\rangle+\mathrm{i}|l\rangle}$ & 
        $F_t$ & 
        $F_{\mathrm{Classical}}$
        \\ 
        
        $99.6(1)\%$ &
        $99.6(1)\%$ & 
        $99.2(1)\%$ &
        $99.1(1)\%$ &
        $99.3(1)\%$ &
        $70.0\%$
        \\
 
    \end{tabular}
        \caption{\label{fidelity} Storage fidelity of time-bin qubits. $F_{|e\rangle}$: measured fidelity for $|e\rangle$.
        $F_{|l\rangle}$: measured fidelity for $|l\rangle$.
        $F_{|e\rangle+|l\rangle}$: measured fidelity for ${|e\rangle+|l\rangle}$.
        $F_{|e\rangle+\mathrm{i}|l\rangle}$: measured fidelity for ${|e\rangle+\mathrm{i}|l\rangle}$.
        $F_t=\frac{1}{3}F_{el}+\frac{2}{3}F_{+-}$, where $F_{el}$ ($F_{+-}$) is the average fidelity of pole (superposition) states.
        $F_{\mathrm{Classical}}$: the maximal fidelity that can be achieved with the classical measure-and-prepare strategy (definitions detailed in Supplementary Materials Section E).
     }
    \end{ruledtabular}

\end{table}

\textit{Conclusion.} In conclusion, we have demonstrated an integrated quantum memory with a storage-and-retrieval efficiency exceeding 90\% for single-photon-level inputs, establishing a new benchmark for on-chip photonic quantum memories. The current device has a volume of $1.2\times10^{-3}~\text{mm}^{3}$, more than three orders of magnitude smaller than that of atomic gas systems achieving similar storage efficiencies~\cite{Wang2019,kbwj-md9n}. Such a small device volume is particularly attractive for the future development of long-lived spin-wave quantum storage, as it could substantially reduce the optical and RF power required for control fields~\cite{10.1093/nsr/nwae161,lv2025minutescalephotonicquantummemory,doi:10.1126/sciadv.adu5264,ortu_storage_2022,PhysRevLett.114.230502}. In addition, the waveguide-cavity architecture enables dense spatial multiplexing and further miniaturization~\cite{OZW} for scalable applications.

%By combining an ultra-low-loss waveguide cavity with spectral engineering and waveform matching that overcome the linewidth limitation in high-optical-depth Pr$^{3+}$ systems, our work identifies a practical route towards high-efficiency and long-lived integrated quantum memories.
\begin{acknowledgments}
P.-X. L., Y.-X. X. and T.-X. Z. contributed equally to this work. This work is supported by the Strategic Priority Research Program of the Chinese Academy of Sciences (No. XDB1740000), HFNL Self-Deployed Project (No. ZB2026020200), the Quantum Science and Technology-National Science and Technology Major Project (No. 2021ZD0301200), the National Natural Science Foundation of China (Nos. 12222411 and 12404572) and this work was partially carried out at the USTC Center for Micro and Nanoscale Research and Fabrication. Z.-Q. Z. acknowledges the support from the Youth Innovation Promotion Association CAS.
\end{acknowledgments}

\textit{Data Availability}--The data that support the findings of this article are openly available~\cite{dataset}.

%\section*{Author contributions}
%Z.-Q.Z. proposed the original protocol and supervised all aspects of this work; P.-X. L. improved the protocol and calculated the entanglement distribution time, Y.-P. L. improved the channel design and calculated the channel efficiency, P.-X. L., Y.-P. L. and Z.-Q.Z. wrote the manuscript with input from others. Z.-Q.Z. and C.-F.L. supervised the project. All authors discussed the calculation procedures and results.\\
 
%%%%%%%%%%%%%%%%%%%%%%%%%%%%%%%%%%%%%%%%%%%%%%%%%%%%%%%%
%\input{output.bbl}
\bibliography{bibliography.bib}
\end{document}